\documentclass[pr,aps,twocolumn,nofootinbib]{revtex4}
\usepackage{amsfonts}
\usepackage{amsmath}
\usepackage{amssymb}
\usepackage{graphicx}
\usepackage{caption}
\providecommand{\U}[1]{\protect\rule{.1in}{.1in}}

\begin{document}
\title{The paradigm of kinematics and dynamics must yield to causal structure}

\author{Robert W. Spekkens}
\affiliation{Perimeter Institute for Theoretical Physics, Waterloo, Ontario, Canada N2L
2Y5}
\date{August 30, 2012}

\begin{abstract}
The distinction between a theory's kinematics and its dynamics, that is, between the space of physical states it posits and its law of evolution, is central to the conceptual framework of many physicists.
A change to the kinematics of a theory, however, can be compensated by a change to its dynamics without empirical consequence, which strongly suggests that these features of the theory, considered separately, cannot have physical significance.  It must therefore be concluded (with apologies to Minkowski) that henceforth kinematics by itself, and dynamics by itself, are doomed to fade away into mere shadows, and only a kind of union of the two will preserve an independent reality.  The notion of causal structure seems to provide a good characterization of this union.
\end{abstract}
\maketitle


Proposals for physical theories generally have two components: the first is a
specification of the space of physical states that are possible according to
the theory, generally called the \emph{kinematics }of the theory, while the
second describes the possibilities for the evolution of the physical state,
called the \emph{dynamics. \ }This distinction is ubiquitous. \ Not only do we
recognize it as a feature of the empirically successful theories of the past,
such as Newtonian mechanics and Maxwell's theory of electromagnetism, it
persists in relativistic and quantum theories as well and is even conspicuous
in proposals for novel physical theories. \ Consider, for instance, some recent
proposals for how to unify quantum theory and gravity. Fay Dowker describes
the idea of causal histories as follows~\cite{dowker2005causal}:
\begin{quote}
The hypothesis that the deep structure of spacetime is a discrete poset
characterises causal set theory at the kinematical level; that is, it is a
proposition about what substance is the subject of the theory. However,
kinematics needs to be completed by dynamics, or rules about how the substance
behaves, if one is to have a complete theory.
\end{quote}
She then proceeds to describe the dynamics. As another example, Carlo
Rovelli describes the basics of loop quantum gravity in the following terms~\cite{rovelli2008loop}:
\begin{quote}
The kinematics of the theory is well understood both physically (quanta of
area and volume, discrete geometry) and from the mathematical point of view.
The part of the theory that is not yet fully under control is the dynamics,
which is determined by the Hamiltonian constraint.
\end{quote}
In the field of quantum foundations, there is a particularly strong insistence
that any well-formed proposal for a physical theory must specify both kinematics and dynamics.
For instance, Sheldon Goldstein describes the deBroglie-Bohm interpretation~\cite{bohm1952suggested} by specifying its kinematics and its dynamics~\cite{GoldsteinSEP}:
\begin{quote}
In Bohmian mechanics a system of particles is described in part by its wave
function, evolving, as usual, according to Schr\"{o}dinger's equation.
However, the wave function provides only a partial description of the system.
This description is completed by the specification of the actual positions of
the particles. The latter evolve according to the ``guiding
equation,'' which expresses the velocities of the particles
in terms of the wave function.
\end{quote}
John Bell provides a similar description of his proposal for a pilot-wave
theory for fermions in his characteristically whimsical style~\cite{bell1986beables}:
\begin{quote}
In the beginning God chose 3-space and 1-time, a Hamiltonian H, and a state
vector $\left\vert 0\right\rangle .$ \ Then She chose a fermion configuration
$n\left(  0\right)  .$ \ This She chose at random from an ensemble of
possibilities with distribution $D\left(  0\right)  $ related to the already
chosen state vector $\left\vert 0\right\rangle .$ \ Then She left the world
alone to evolve according to [the Schr\"{o}dinger equation] and [a stochastic
jump equation for the fermion configuration].
\end{quote}
The distinction persists in the Everett interpretation~\cite{everett1957relative}, where the set of
possible physical states is just the set of pure quantum states, and the
dynamics is simply given by Schr\"{o}dinger's equation (the appearance of
collapses is taken to be a subjective illusion). It is also present in
dynamical collapse theories~\cite{ghirardi1986unified,ghirardi1990markov}, where the kinematics is often
taken to be the same as in Everett's approach --- nothing but wavefunction ---
while the dynamics is given by a stochastic equation that is designed to yield
a good approximation to Schr\"{o}dinger dynamics for microscopic systems and
to the von Neumann projection postulate for macroscopic systems.

While proponents of different interpretations of quantum
theory and proponents of different approaches to quantizing gravity may disagree about the correct kinematics and dynamics, they typically agree that any proposal must be described in these terms.

In this essay, I will argue that the distinction is, in fact, conventional: kinematics and dynamics only
have physical significance when considered jointly, not separately.


In essence, I adopt the following methodological principle: any difference between two physical models that does not yield a difference at the level of empirical phenomena does not correspond to a physical difference and should be eliminated.  Such a principle was arguably endorsed by Einstein when, from the empirical indistinguishability of inertial motion in free space on the one hand and free-fall in a gravitational field on the other, he inferred that one must reject any model which posits a physical difference between these two scenarios (the strong equivalence principle).

Such a principle does not force us to operationalism, the view that one should only seek to make claims about the outcomes of experiments.  For instance, if one didn't already know that the choice of gauge in classical electrodynamics made no difference to its empirical predictions, then discovery of this fact would, by the lights of the principle, lead one to renounce real status for the vector potential in favour of only the electric and magnetic field strengths.  It would not, however, justify a blanket rejection of \emph{any} form of microscopic reality.

As another example, consider the prisoners in Plato's cave who live out their lives seeing objects only through the shadows they cast.  Suppose one of the prisoners strikes upon the idea that there is a third dimension, that objects have a three-dimensional shape, and that the patterns they see are just two-dimensional projections of this shape.  She has constructed a hidden variable model for the phenomena.  Suppose a second prisoner suggests a different hidden variable model, where in addition to the shape, each object has a property called colour which is completely irrelevant to the shadow that it casts.  The methodological principle dictates that because the colour property can be varied without empirical consequence, it must be rejected as unphysical. The shape, on the other hand, has explanatory power and the principle finds no fault with it.  Operationalism, of course, would not even entertain the possibility of such hidden variables.

The principle tells us to constrain our model-building in such a way that every aspect of the posited reality has some explanatory function.  If one takes the view that part of achieving an adequate explanation of a phenomenon is being able to make predictions about the outcomes of interventions and the truths of counterfactuals, then what one is seeking is a \emph{causal} account of the phenomenon.  This suggests that the framework that should replace kinematics and dynamics is one that focuses on causal structure.  I will, in fact, conclude with some arguments in favour of this approach.

\section*{Different formulations of classical mechanics}

Already in classical physics there is ambiguity about how to make the separation
between kinematics and dynamics. \ In what one might call the \emph{Newtonian} formulation of classical
mechanics, the kinematics is given by configuration space, while in the
\emph{Hamiltonian} formulation, it is given by phase space, which considers the
canonical momentum for every independent coordinate to be on an equal footing with the coordinate. For instance, for a single particle, the kinematics of the Newtonian formulation is the space of possible positions while that of the Hamiltonian formulation is the space of possible pairs of values of position and momentum.  The two formulations are still able to make the same empirical predictions because they posit different dynamics. In the Newtonian approach, motion is governed by the Euler-Lagrange equations which are
second-order in time, while in the Hamiltonian approach, it is governed
by Hamilton's equations which are first order in time.

So we can change the kinematics from configuration space to phase space and maintain the same empirical predictions
by adjusting the dynamics accordingly. It's not possible to determine which
kinematics, Newtonian or Hamiltonian, is the \emph{correct }kinematics.   Nor
can we determine the correct dynamics in isolation. The kinematics and dynamics of a theory can only ever be subjected to experimental trial as a pair.

\section*{On the possibility of violating unitarity in quantum dynamics}

Many researchers have suggested that the correct theory of nature might be one
that shares the kinematics of standard quantum theory, but which posits a
different dynamics, one that is not represented by a unitary operator.  There have been many different motivations for considering this possibility. Dynamical collapse theorists, for instance, seek to relieve the tension between a system's free evolution and its evolution due to a measurement.  Others have been motivated to resolve the black hole information loss paradox.  Still others have proposed such theories simply as foils against which the predictions of quantum theory can be tested~\cite{weinberg1989testing}.

Most of these proposals posit a dynamics which is linear in the quantum state (more precisely, in the density
operator representing the state). For instance, this is true of the prominent examples of
dynamical collapse models, such as the proposal of Ghirardi, Rimini and Weber~\cite{ghirardi1986unified}
and the continuous spontaneous localization model~\cite{ghirardi1990markov}.
This linearity is not an incidental feature of these models. Most theories which posit dynamics that are nonlinear also allow superluminal signalling, in contradiction with relativity theory~\cite{gisin1990weinberg}. Such nonlinearity can also lead to trouble with the second law of thermodynamics~\cite{peres1989nonlinear}.

There is an important theorem about linear dynamics that is critical for our analysis: such dynamics can always be understood to arise by adjoining to the system of interest an auxiliary system prepared in some fixed quantum state, implementing a unitary evolution on the composite, and finally throwing away or ignoring the auxiliary system. This is known as the Stinespring dilation theorem~\cite{stinespring1955positive} and is well-known to quantum information theorists\footnote{It is analogous to the fact that one can simulate indeterministic dynamics on a system by deterministic dynamics which couples the system to an additional degree of freedom that is probabilistically distributed.}.

All proposals for nonunitary but linear modifications of quantum theory presume that it is in fact possible to distinguish the predictions of these theories from those of standard quantum mechanics. \ For instance, the
experimental evidence that is championed as the ``smoking gun'' which would rule in favour of such a modification
is \emph{anomalous decoherence }--- an increase in the entropy of the quantum state of a system which cannot be accounted for by an interaction with the system's environment. Everyone admits that such a signature is
extremely difficult to detect if it exists. But the point I'd like to make here is that
\emph{even if} such anomalous decoherence were detected, it would not vindicate
the conclusion that the dynamics is nonunitary. Because of the Stinespring dilation
theorem, such decoherence is also consistent with the
assumption that there are some hitherto-unrecognized degrees of freedom and that the quantum system under investigation is coupled unitarily to these\footnote{A collapse theorist will no doubt reject this explanation on the grounds that one cannot solve the quantum measurement problem while maintaining unitarity. Nonetheless, our argument shows that someone who does not share their views on the quantum measurement problem need not be persuaded of a failure of unitarity.}.

So, while it is typically assumed that such an anomaly would reveal that quantum theory was mistaken in its \emph{dynamics}, we could just as well take it to reveal that quantum theory was correct in its dynamics but mistaken in its \emph{kinematics}. The experimental evidence alone cannot decide
the issue.  By the lights of our methodological principle, it follows that the distinction must be purely conventional.


\section*{Freedom in the choice of kinematics for pilot-wave theories}


The pilot-wave theory of deBroglie and Bohm supplements the wavefunction with additional variables, but it turns out that there is a great deal of freedom in how to choose these variables.  A simple example of this arises for the case of spin.
Bohm, Schiller, and Tiomno have proposed that particles with spin should be modeled as
extended rigid objects and that the spinor wavefunction should be supplemented
not only with the positions of the particles (as is standardly done for particles without spin), but with their orientation in space as well~\cite{bohm1955causal}.  In addition to the equation which governs the
evolution of the spinor wavefunction (the Pauli equation), they propose a
guidance equation that specifies how the positions and orientations evolve over time.

But there is another, more minimalist, proposal for how to deal with spin, due to Bell~\cite{bell1982impossible}. The only variables that supplement the wavefunction in his approach are the particle
positions.  The theory nonetheless makes the same predictions as the
one without spin
because the equations of motion for the particle positions depend on the spinor wavefunction.
These two approaches make exactly the same experimental predictions.  This is possible because our experience of quantum phenomena consists of observations of macroscopic variables such as pointer positions rather than direct observation of the properties of the particle.


The non-uniqueness of the choice of kinematics for pilot-wave theories is not isolated to spin.  It is generic.
The case of quantum electrodynamics (QED) illustrates this well. Not only is there
a pilot-wave theory for QED, there are multiple viable proposals, all of which
produce the same empirical predictions. You could follow Bohm's treatment of
the electromagnetic field, where the quantum state is supplemented by the
configuration of the electric field~\cite{bohm1952suggestedII}. Alternatively, you could make the
supplementary variable the magnetic field, or any other linear combination of the two.
For the charges, you could use Bell's discrete model of fermions on a
lattice (mentioned in the introduction), where the supplementary variables are
the fermion numbers at every lattice point~\cite{bell1986beables}. Or, if you preferred, you could
use Colin's continuum version of this model\cite{colin2003deterministic}. If you fancy something a
bit more exotic, you might prefer to adopt Struyve and Westman's minimalist
pilot-wave theory for QED, which treats charges in a manner akin to how Bell
treats spin~\cite{struyve2007minimalist}.  Here, the variables that are taken to supplement the quantum states are
\emph{just} the electric field strengths.
\ No variables for the charges are introduced.  By virtue of
Gauss's law, the field nonetheless carries an image of all the charges and hence it
carries an image of the pointer positions. \ This image is what we infer when our
eyes detect the fields.  But the charges are an illusion.  And, of course, according to this model the stuff of which we are made is not charges either: we are fields observing fields.

The existence of many empirically adequate versions of Bohmian mechanics has
led many commentators to appeal to principles of simplicity or elegance to try
to decide among them. An alternative response is suggested by our
methodological principle: any feature of the theory that varies among the
different versions is not physical.

\section*{Kinematical locality and dynamical locality}

I consider one final example, the one that first set me down the
path of doubting the significance of the distinction between kinematics and
dynamics. \ It concerns different notions of locality within \emph{ontological models} of quantum theory.  Such models posit that systems are described by properties, the complete specification of which is called the \emph{ontic state} of the system, and that measurements reveal information about those properties~\cite{spekkens2005contextuality}.

It is natural to say that an ontological model has \emph{kinematical locality} if, for any two
systems $A$ and $B$, every ontic state $\lambda_{AB}$ of the composite is
simply a specification of the ontic state of each component,
\[
\lambda_{AB}=\left(  \lambda_{A},\lambda_{B}\right)  .
\]
In such a theory, once you have specified all the properties of $A$ and of $B,$ you have specified all of the properties of the composite $AB$.  In other words, kinematical locality says that there are
no holistic properties \footnote{The assumption has also been called \emph{separability}~\cite{harrigan2010einstein}.}.

It is also natural to define a dynamical notion of locality for
relativistic theories: a change to the ontic state $\lambda_{S}$ of a
localized system $S$ cannot be a result of a change to the ontic state
$\lambda_{S^{\prime}}$ of a localized system $S^{\prime}$ if $S^{\prime}$ is
outside the backward light-cone of $S.$ In other words, against the backdrop of
a relativistic space-time, this notion of locality asserts that all causal
influences propagate at speeds that are no faster than the speed of light.

Note that this definition of dynamical locality
has made reference to the ontic state $\lambda_{S}$ of a
\emph{localized} system $S.$ \ If $S$ is part of a composite system with holistic properties, then the ontic state
of this composite, $\lambda_{SS^{\prime\prime}}$, need not factorize into $\lambda_{S}$ and
$\lambda_{S^{\prime\prime}}$ therefore we cannot necessarily even
define $\lambda_{S}.$ \ In this sense, the dynamical notion of locality already presumes the kinematical one.

It is possible to derive Bell inequalities starting from these assumptions of
locality (and a few other assumptions such as the freedom of measurement
settings and the absence of retrocausal influences).  Famously, quantum
theory predicts a violation of the Bell inequalities.  In the face of this
violation, one must give up one or more of the assumptions. Locality is a
prime candidate to consider and if we do so, then the following question naturally arises: is it possible to accommodate violations of Bell inequalities by admitting a failure of the dynamical notion of locality
while salvaging the kinematical notion?

It turns out that for any realist interpretations of quantum theory wherein the ontic state encodes the quantum state ($\psi$-ontic models in the terminology of Ref.~\cite{harrigan2010einstein})), there is a failure of \emph{both} sorts of locality.  In such models, kinematical locality fails simply by virtue of the existence of entangled states.  This is the case for all of the interpretations enumerated in the introduction: Everett, collapse theories, and deBroglie-Bohm.
Might there nonetheless be some alternative to these interpretations that \emph{does} manage to salvage kinematical locality?

I've told the story in such a way that this seems to be a perfectly meaningful
question. But I would like to argue that, in fact, it is not.

To see this, it suffices to realize that it is \emph{trivial }to build a model of quantum theory
that salvages kinematical locality. For example, we can do so by a
slight modification of the deBroglie-Bohm model.  Because the
particle positions can be specified locally, the only obstacle to satisfying kinematical locality is that the other part of the ontology, the universal wavefunction, does not factorize across systems
and thus must describe a holistic property of the universe.  This conclusion, however, relied on a particular way of associating the wavefunction with space-time.  Can we imagine a different association that would make the model kinematically local?  Sure.  Just put a copy of the universal wavefunction at every point in space.  It can then pilot the motion of every particle by a local influence. Alternatively, you could put it at the location of the center of mass of the universe and have it achieve its piloting by superluminal influence --- remember, we are
allowing arbitrary violations of dynamical locality.  Or, put it under the corner of my doormat and let it choreograph all of the particles in the universe from there.

The point is that the failure of dynamical locality yields so much leeway in the
dynamics that one can easily accommodate any sort of kinematics, including a
local kinematics. Of course, these models are not credible and no one would seriously propose them
\footnote{Norsen has proposed a slightly more credible model but only as a proof of principle that kinematical locality can indeed be achieved\cite{norsen2010theory}.}, but what this suggests to me is \emph{not} that we should look for nicer models, but rather that the question of whether one can salvage kinematical locality was not an interesting one after all.  The mistake, I believe, was to take seriously the distinction between kinematics and dynamics.

\section*{Summary of the argument}

A clear pattern has emerged. \ In all of the examples considered, we seem to be able to
accommodate wildly different choices of kinematics in our models without changing their empirical predictions simply by modifying the dynamics, and vice-versa.
This strikes me as strong evidence for the view that the distinction between kinematics
and dynamics --- a distinction that is often central to the way that physicists characterize their best theories and to the way they constrain their theory-building --- is purely conventional and should be abandoned.

\section*{From kinematics and dynamics to causal structure}

Although it is not entirely clear at this stage what survives the elimination of the
distinction between kinematics and dynamics, I would like to suggest a
promising candidate: the concept of \emph{causal structure.}

\begin{figure} [h]
\begin{center}
\includegraphics[scale=0.6]{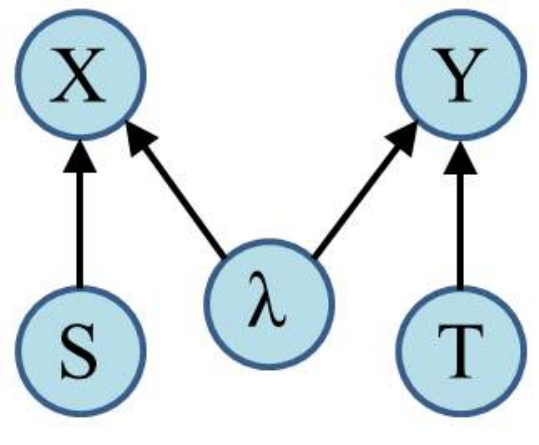}
\caption{The causal graph associated with Bell's notion of local causality}\label{fig:LocalCausality}
\end{center}
\end{figure}

\begin{figure} [h]
\begin{center}
\includegraphics[scale=0.6]{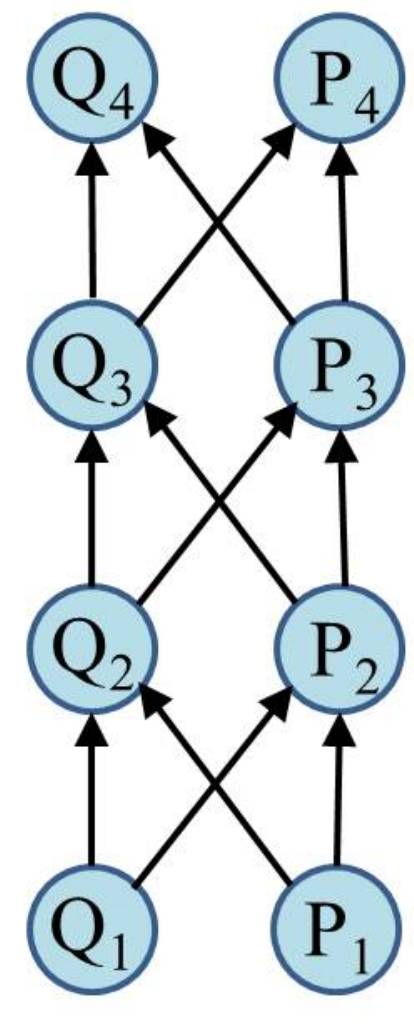}
\quad \quad \quad
\includegraphics[scale=0.6]{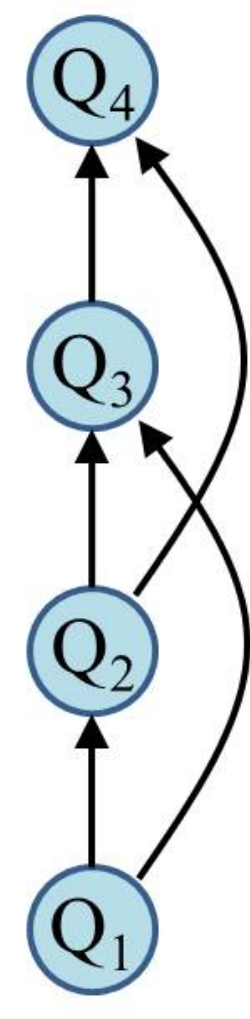}
\caption{Causal graphs for Hamiltonian and Newtonian formulations of mechanics respectively.}\label{fig:HL}
\end{center}
\end{figure}

In recent years, there has been significant progress in providing a rigorous mathematical formalism for expressing causal relations and for making inferences from these about the consequences of interventions and the truths of counterfactuals.  The work has been done primarily by researchers in the fields of statistics, machine learning, and philosophy and is well summarized in the texts by Spirtes, Glymour, and Scheines~\cite{Spirtes2001}, and Pearl~\cite{Pearl2009}.  According to this approach, the causal influences that hold among a set of classical variables can be modeled by the arrows in a directed acyclic graph, of the sort depicted in
Figs.~\ref{fig:LocalCausality} and \ref{fig:HL}, together with some causal-statistical parameters describing the
strengths of the influences.  If $\mathrm{Parents}\left(  X\right)$ denote
the variables that are direct causes, i.e. causal parents, of a variable $X$,
then the causal-statistical parameters are conditional probabilities $P\left(
X|\mathrm{Parents}\left(  X\right)  \right)  $ for every $X$.  If a variable $X$ has no parents within the model, then one simply specifies $P(X)$. The graph and the parameters together constitute the causal model.

It remains only to see why this framework has some hope of capturing the nonconventional elements of the various examples I have presented.

The strongest argument in favour of this framework is that it provides a way to move beyond kinematical and dynamical notions of locality.  John Bell was someone who clearly endorsed the kinematical-dynamical paradigm of model-building, as the quote in the introduction illustrates, and who recognized the distinction among notions of locality, referring to models satisfying kinematical locality as theories of ``local beables''~\cite{bell1976theory}.
In his most precise formulation of the notion of locality, however, which he called local \emph{causality}, he appears to have transcended the paradigm of kinematics and dynamics and made an early foray into the new paradigm of causal structure.

Consider a Bell-type experiment. A pair of systems, labeled $A$ and $B,$ are prepared together
and then taken to distant locations. The variable that specifies the choice of
measurement on $A$ (respectively $B$) is denoted $S ($respectively $T)$ and the
variable specifying the measurement's outcome is denoted $X ($respectively
$Y).$ \ Bell interprets the question of whether a set of correlations
$P\left(  XY|ST\right)  $ admits of a locally causal explanation as the
question of whether the correlations between $X$ and $Y$ can be entirely
explained by a common cause $\lambda,$ that is, whether they can be explained
by a causal graph of the form illustrated in Fig.~\ref{fig:LocalCausality}.  From the causal
model, we derive that
\[
P\left(  XY|ST\right)  =\sum_{\lambda}P\left(  X|S\lambda\right)  P\left(
Y|T\lambda\right)  P(\lambda),
\]
Correlations $P(XY|ST)$ of this form can be shown to satisfy certain inequalities, called the Bell inequalities, which can be tested by experiments (and are found to be violated in a quantum world).

If we think of the variable $\lambda$ as the ontic state of the composite $AB,$ then
we see that we have not needed to specify whether or not $\lambda$ factorizes
as $\left(  \lambda_{A},\lambda_{B}\right).$ \ Bell recognized this fact and emphasized it in his later writing: ``It is notable that in this argument nothing is said about the locality, or even localizability, of the
variable $\lambda$~\cite{bell1981bertlmann}.''
In other words, Bell managed to make empirical claims about a class of ontological models without needing to
make any commitments about the separate nature of the kinematics and the dynamics
of those models!  I suggest that this approach should be considered a template for future physics.

It is not as clear how the paradigm of causal structure overcomes the conventionality of the kinematics-dynamics distinction in the other examples I've presented, but there are some interesting clues that this is the right track.

Consider the example of Hamiltonian and Newtonian formulations of mechanics. If we let $Q_{i}$ denote a coordinate at time $t_{i}$ and $P_{i}$ its canonically conjugate momentum, then the causal models associated respectively with the two approaches are depicted in Fig.~\ref{fig:HL}. The fact that Hamiltonian dynamics is
first-order in time implies that the $Q$ and $P$ variables at a given
time are causally influenced directly only by the $Q$ and $P$ at the previous
time. \ Meanwhile, the second-order nature of Newtonian dynamics is
captured by the fact that $Q$ at a given time is causally influenced directly
by the $Q$s at \emph{two} previous times. \ In both models, we have a
causal influence from $Q_{1}$ to $Q_{3}$, but in the Newtonian
case it is direct, while in the Hamiltonian case it is mediated by $P_{2}.$
Nonetheless, the kinds of correlations that can be made to hold between $Q_{1}$ and $Q_{3}$ are the same regardless of whether the causal influence is direct or mediated by $P_2$\footnote{There is a subtlety here: it follows from the form of the causal graph in the Newtonian model that $Q_{1}$ and $Q_{4}$ are conditionally independent given $Q_{2}$ and $Q_{3}$, but in the Hamiltonian case, this fact must be inferred from the causal-statistical parameters.}.  The consequences for $Q_3$ of interventions upon the value of $Q_1$ also are insensitive to this difference.
So from the perspective of the paradigm of causal structure, the Hamiltonian and Newtonian formulations appear less distinct than they do if one focusses on kinematics and dynamics.

Empirical predictions of statistical theories are typically expressed in terms of statistical dependences among variables that are observed or controlled.  My guiding methodological principle suggests that we should confine our attention to those causal features that are relevant for such dependences.  In other words, although we can convert a particular claim about kinematics and dynamics into a causal graph, not all features of this graph will have relevance for statistical dependences. Recent work that seeks to infer causal structure from observed correlations has naturally gravitated towards the notion of equivalence classes of causal graphs, where the equivalence relation is the ability to produce the same set of correlations.  One could also try to characterize equivalence classes of causal models while allowing for restrictions on the forms of the conditional probabilities or when one allows not only \emph{observations} of variables but \emph{interventions} upon them.  Such equivalence classes, or something like them, seem to be the best candidates for the mathematical objects in terms of which our classical models should be described.

Finally, by replacing conditional probabilities with quantum operations, one can define a quantum generalization of causal models ---\emph{quantum causal models}~\cite{Leifer2006,Leifer2011} --- which appear promising for providing a realist interpretation of quantum theory. It is equivalence classes of causal structures here that are likely to provide the best framework for future physics.

The paradigm of kinematics and dynamics has served us well. So well, in fact, that it is woven deeply into the fabric of our thinking about physical theories and will not be easily supplanted. I have nonetheless argued that we must abandon it.  Meanwhile, the paradigm of causal structure is nascent, unfamiliar and incomplete, but it seems ideally suited to capturing the nonconventional distillate of the union of kinematics and dynamics and it can already claim an impressive achievement in the form of Bell's notion of local causality.

Rest in peace kinematics and dynamics.  Long live causal structure!

\bibliographystyle{unsrt}						
\bibliography{FQXIessay-Ref2}

\end{document}